\documentstyle[aps,preprint,epsfig]{revtex}

\begin{document}

\draft

\def\DD{{\Delta \Delta}}
\def\DN{{\Delta N}}
\def\ND{{N \Delta}}

\title{Double $\Delta(1232)$ excitation and the ABC effect \\ 
  in the reaction $n+p \to ^2$H$\, (\pi\pi)$}

\author{C.~A. Mosbacher and F.Osterfeld$^{\,\dagger}$}

\address{Institut f\"ur Kernphysik, Forschungszentrum J\"ulich GmbH, 
  D--52425 J\"ulich, Germany}

\date{\today}

\maketitle


\begin{abstract}

The deuteron spectrum in $n+p \to ^2$H$\, (\pi\pi)$
at $k_n = 1.88$ GeV/c and $\theta_d = 0^o$
is explained by considering a $\Delta \Delta$ excitation as the dominant reaction  
mechanism for the $2\pi$ production. 
We present a new theoretical approach based on a coupled channel 
formalism which allows to include the residual interaction
within the intermediate $\DD$ and $\DN$ systems. The corresponding
interaction potentials $V_\DD$ and $V_\DN$ are adopted from a meson
exchange model with $\pi$, $\rho$, $\omega$, and $\sigma$ exchange
taken into account.
The influence of the residual interaction on the deuteron spectrum is studied.
We also predict the angular distribution of the two pions. It is shown that this
distribution is closely connected to the spin structure of the $\DD$
excitation.

\end{abstract}
                                                                                 



\section{Introduction}

The ABC effect, a mesonic structure of isospin $I=0$, was first observed by
Abashian, Booth and Crow in the reaction $p+^2H \to ^3$He$ + (\pi\pi)$
\cite{abashian61,abashian63,banaigs73}. 
Later on, it was also shown to appear in several other hadronic reactions,
like e.g. in $n+p \to ^2$H$\, + X$ \cite{plouin78,abdivaliev80}
and in $^2$H$+ ^2$H$\to ^4$He$\, + X$ \cite{banaigs76,willis97}.
In all these cases, the ABC structure shows up as a cross section peak
at missing masses of about 300 -- 350 MeV, and with a position and width
that varies quite rapidly according to the kinematic conditions.
The only possible explanation for the effect seems to be an enhancement 
in the double pion production which is caused by a strongly energy 
dependent production amplitude.

Since the ABC effect is present in 
reactions of different type, it seems to be quite natural to assume 
that there exists a common underlying mechanism for the $2\pi$ production
which should be able to explain all the different experiments.
From a theoretical point of view, $n+p \to ^2$H$\, (\pi \pi)$ has 
to be the reaction to analyze first since it involves only two-nucleon 
states beside the pions and can thus be treated in the most accurate and direct way.  
For this reaction, the ABC enhancement corresponds to a beam momentum of 
approximately 2 GeV/c in the laboratory system and to a center of mass
energy of $\sqrt{s} \approx 2 M_\Delta$, i.e. twice the mass of the
$\Delta (1232)$ resonance. This fact strongly suggests 
that the two pions are dominantly produced via a double $\Delta$ 
excitation. Indeed, as it has already been shown by Bar--Nir et al.\ 
\cite{barnir75,risser73}, the main features of the experimental spectrum
can be understood by assuming the $\DD$ mechanism to be the most 
efficient one for $2 \pi$ production in the energy region under consideration. 

In the present work, we will present a new calculation of $\DD$ and $\DN$
excitations
in the $n+p \to ^2$H$\, (\pi \pi)$ reaction which explicitly includes the 
effects of residual interactions in the intermediate resonance states
\cite{mosbacher98b,mosbacher98c}. 
Our aim is to show that the $2\pi$ production may serve as 
an interesting tool for further examination of direct $\DD$ and $\DN$ potentials 
which are not very well known. For this purpose, we will also examine how the
spin structure of the $\DD$ excitation influences the angular distribution
of the two pions. Our model is based on a coupled channel approach
in a non--relativistic framework. The potentials of the residual interactions 
are adopted from a meson exchange model \cite{mosbacher97,machleidt87}.
In this work, contributions from $\pi$, $\rho$, $\omega$, and $\sigma$
exchange are included. 
The $\Delta$ resonance is treated thereby as a quasi--particle with a given
mass and an energy--dependent, intrinsic width.

Experimental results for the $n+p \to ^2$H$\, (\pi\pi)$ cross section 
in the ABC energy region have been 
obtained by Plouin et al.\ \cite{plouin78} at $k_n = 1.88$ GeV.
In order to allow for a direct comparison with our theoretical results,
the contribution from one pion production has been subtracted from the spectra.
It has been stressed that there might be some additional background 
from $\eta$ production \cite{plouin90} due to momentum spread of the beam, but
since there is no microscopic model for this contribution available,
we decided not to perform any further corrections.
One should keep in mind, however, that not all of the 
uncertainties have been included in the experimental error bars.

In section \ref{sec:1}, we will present the theoretical framework of our model.
The results of our calculation are presented and discussed
in section \ref{sec:2}. The paper concludes with a summary in section \ref{sec:3}.


\section{Theoretical framework}
\label{sec:1}

\subsection{Cross section}

We are interested here in the calculation of the deuteron spectrum for 
the $n+p \to ^2$H$\, (\pi\pi)$ reaction, i.e. in the double differential
cross section $d^2 \sigma / dk_d d\Omega_d$. Using relativistic kinematics,
the cross section is given as
\begin{equation}
  \label{eq:1}
  d \sigma = \frac{2 E_p E_n}{\sqrt{\lambda(s,M_p,M_n)}} \: \frac{M_n}{E_n} \:
  \frac{M_p}{E_p} \: \frac{d^3 k_d}{(2 \pi)^3} \; 
  \frac{d^3 k}{(2\pi)^3 (2E_a)(2 E_b)} \; \delta(E_p + E_n - E_d - E_{\pi\pi})
  \, \overline{\bigl| \: {\mathcal M} \: \bigr| ^{2}}
\end{equation}
where $\vec k = \frac{1}{2} (\vec k_a - \vec k_b \,)$ denotes the relative
momentum of the two pions and $E_{\pi\pi} = E_a + E_b$ is the sum of their energies. 
The indices $n(p)$ and $d$ refer to the neutron projectile (proton target)
and the deuteron ejectile, respectively. The center of mass momentum of 
the $2\pi$ system $\vec K = \vec k_a + \vec k_b$ is fixed by momentum
conservation and equals $\vec k_n + \vec k_p - \vec k_d$. 
As usual, the function $\lambda$ is defined to be $\lambda(s,M_p^2,M_n^2) = 
\left[ s-(M_p+M_n)^2 \right] \left[ s-(M_p-M_n)^2 \right]$.
Finally, $| \: {\mathcal M} \: | ^{2}$ denotes the transition amplitude 
for the reaction; 
in the case of the unpolarized cross section, an average over the initial 
and a sum over the final spin orientations has to be performed.

From eq.\ (\ref{eq:1}) one easily obtains the deuteron spectrum
\begin{equation}
  \label{eq:2}
  \frac{d^2 \sigma}{dk_d d\Omega_d} = \frac{1}{(2 \pi)^5} \: \frac{M_n}{k_n^{\text{lab}}}
  \: k_d^2 \: \int \frac{k^2}{4 E_a E_b} \; \frac{dk}{d E_{\pi\pi}} \: d\Omega \; 
  \overline{\bigl| \: {\mathcal M} \: \bigr| ^{2}}
\end{equation}
Since we are going to calculate the transititon matrixelement in a 
non--relativistic approach, the above expression has to be evaluated
in the deuteron rest frame (drf) where the deuteron is well described by 
the usual non--relativistic wave function \cite{machleidt87}. 
Afterwards, the transformation 
of the spectrum into the laboratory frame (lab) is performed by using the
simple relation
\begin{equation}
  \label{eq:3}
  \left( \frac{d^2 \sigma}{dk_d d\Omega_d} \right)_{\text{lab}}
  = \frac{E_d^{\text{drf}} (k_d^{\text{lab}})^2}{E_d^{\text{lab}} 
  (k_d^{\text{drf}})^2}
  \: \left( \frac{d^2 \sigma}{dk_d d\Omega_d} \right)_{\text{drf}} \:.
\end{equation}

\subsection{Reaction mechanism and evaluation of the transition amplitude} 

\begin{figure}
  \begin{center} 
    \epsfig{file=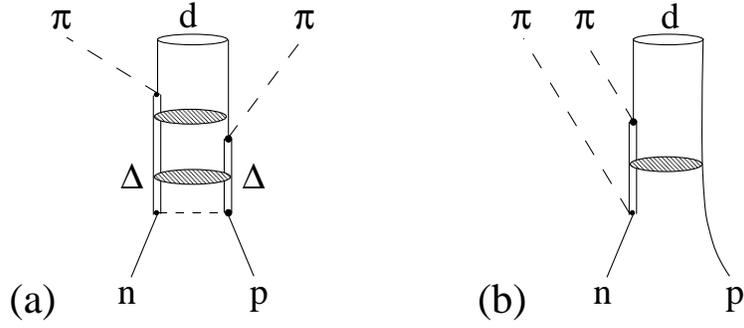, height=4.5cm}
  \end{center}
  \caption{Feynman diagrams of $2 \pi$ production via $\Delta$ excitations
    in the $n+p \to ^2$H$\, (\pi\pi)$ reaction: (a) $\DD$ excitation, 
    (b) $\DN$ excitation. The shaded areas symbolize the residual interaction 
    in the intermediate $\Delta \Delta$ and $\Delta N$ states, respectively.}
  \label{fig:diagrams}
\end{figure}

In our microscopic model for the $n+p \to ^2$H$\, (\pi\pi)$ reaction, 
we take into account
two different reaction mechanisms corresponding to the Feynman diagrams 
presented in fig.\ \ref{fig:diagrams}. The first mechanism (a) involves the 
excitation of a double Delta ($\DD$) intermediate state where both $\Delta$
resonances subsequently decay into a pion and a nucleon.
In the case of the second mechanism (b), each of the two pions couples to
the same $\Delta$, hence only one resonance (i.e. a $\DN$ system) is excited.

For the explicit calculation of the two matrix elements, wa take advantage of 
the principle of detailed balance. This allows us to assume that the deuteron 
is in the initial state and absorbes the two pions, which finally will lead
to a $n+p$ state. The absorption of only one pion with momentum 
$\vec k_\pi$ is described by the operator
\begin{equation}
  \label{eq:4}
  F^{\dagger}_\pi (\vec k_\pi) =  
  e^{i \vec k_\pi \cdot \vec r /2} \: \frac{f_{\pi N \Delta}}{m_\pi}
  \; ( \vec S_1^\dagger \cdot \vec k_\pi \,) \; T_1^\dagger \; + (1 \leftrightarrow 2) \, .
\end{equation}
The exponential term represents the plane wave of the pion field. The 
structure of the $\pi \DN$ coupling follows from the usual interaction
lagrangian \cite{machleidt87} in the non--relativistic reduction.
$\vec S^\dagger$ ($\vec T^\dagger$) is the spin (isospin) transition matrix 
\cite{weber78} for the $\Delta$ excitation, and $\vec r = \vec r_1 - \vec r_2$ 
is the relative coordinate of the two nucleons in the deuteron. 
The symbol $(1 \leftrightarrow 2)$ in eq.\ (\ref{eq:4}) shall indicate
that the operator has to be properly symmetrized with respect to the
nucleonic coordinates, i.e. each of the two nucleons 
in the deuteron may be excited to a $\Delta$ resonance. 

\begin{figure}
  \begin{center} 
    \epsfig{file=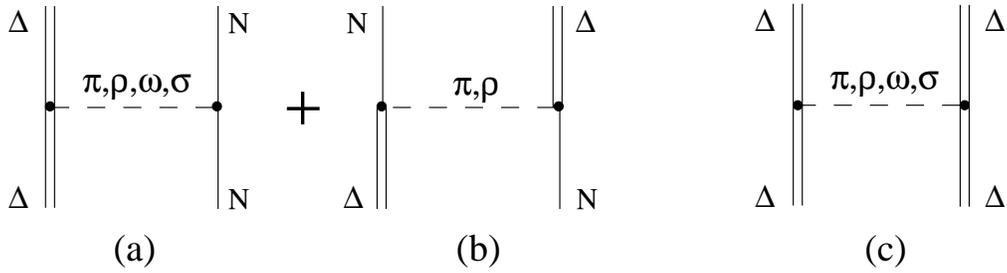, height=3.6cm}
  \end{center}
  \caption{Residual interactions in the meson exchange picture.
    (a) Direct and (b) exchange contribution of the $\DN$ potential,
    (c) $\DD$ potential. The mesons taken into account are
    $\pi$, $\rho$, $\omega$, and $\sigma$.}
  \label{fig:potential}
\end{figure}

Following refs.\ \cite{udagawa94,mosbacher97}, the quantum mechanical state
\begin{equation}
  \label{eq:5}
  | \rho_\DN \rangle = F^{\dagger}_\pi (\vec k_\pi) \: | \psi_d \rangle 
\end{equation}
is called the uncorrelated source function for the $\DN$ intermediate system,
with $| \psi_d \rangle$ being the deuteron wave function.
After propagation of the $\DN$ system and absorption of a second pion, one 
obtains the source function for the $\DD$ state,
\begin{equation} 
  \label{eq:6}
  | \rho_\DD \rangle = 
  F^{\dagger}_\pi (\vec k_b) \: G_\DN \: 
  F^{\dagger}_\pi (\vec k_a) \: | \psi_d \rangle .
\end{equation}
In our model, the full propagator $G_\DN$ is given by 
\begin{equation}
  \label{eq:7}
  G_\DN = \frac{1}{ \epsilon_\DN + \frac{i}{2} \Gamma_\Delta (s_\Delta)
  - T_\DN - V_\DN } \:.
\end{equation}
It contains the excitation energy $\epsilon_\DN$,
the energy--dependent width $\Gamma_\Delta (s_\Delta)$ of the resonance
\cite{dmitriev86}, the operator of the kinetic energy $T_\DN$ and the 
interaction potential $V_\DN$ of the intermediate $\DN$ system. 
$\epsilon_\DN$ and $s_\Delta$ are fixed by imposing energy conservation 
at the $\pi \ND$ vertex.
The potential $V_\DN$ is constructed in a meson exchange model with
$\pi$, $\rho$, $\omega$ and $\sigma$ exchange taken into account.
A diagrammatic representation of the residual $\DN$ interaction is given in
fig.\ \ref{fig:potential} (a,b). 
For more details, we refer the reader to refs.\ \cite{mosbacher97,mosbacher98c}
where also explicit expressions for $V_{\DN}$ can be found. 

In eq.\ (\ref{eq:6}) we have not yet considered that the $\DD$
source function has to be symmetric under exchange of the pions $a$ and $b$.
According to the value $T$ of total isospin in the $\pi\pi$ system, 
we define the symmetrized operator
\begin{eqnarray}
  \label{eq:8}
  F_{\pi\pi}^\dagger (\vec k_a, \vec k_b) = \frac{1}{\sqrt{2}} \: \left\{
  F_\pi^{\dagger} (\vec k_b) \: G_\DN \: F_\pi^{\dagger} (\vec k_a)
  \pm F_\pi^{\dagger} (\vec k_a) \: G_\DN \: F_\pi^{\dagger} (\vec k_b) \right\} \\
  \nonumber 
  \text{with} \: \left\{ 
  \begin{array}{ll} 
    + \;\; \text{if} & \;\; | T M_T \rangle = | 00 \rangle
        = \frac{1}{\sqrt{3}} | \pi^+ \pi^- - \pi^0\pi^0 + \pi^- \pi^+ \rangle \\ 
    - \;\; \text{if} & \;\; | T M_T \rangle = | 10 \rangle
        = \frac{1}{\sqrt{2}} | \pi^+ \pi^- - \pi^- \pi^+ \rangle
  \end{array} \right. \: .
\end{eqnarray}
The complete matrix element for the $\DD$ mechanism can now be written as
\begin{equation}
  \label{eq:9}
  {\mathcal M}_\DD = \: \langle \psi_n | \: \langle \psi_p | 
  \: V_{\DD \to NN} \, G_\DD \, F^\dagger_{\pi\pi} (\vec k_a, \vec k_b \,) 
  \: | \psi_d \rangle
\end{equation}
Here, $\psi_{n(p)}$ denote the distorted waves 
of the projectile neutron (target proton) which are calculated in 
eikonal approximation.
In analogy to eq.\ (\ref{eq:7}), the full propagator for the $\DD$ system
is given by
\begin{equation}
  G_\DD = \frac{1}{ \epsilon_\DD + \frac{i}{2} \left[ \Gamma_{\Delta_a} (s_{\Delta_a})
    + \Gamma_{\Delta_b} (s_{\Delta_b}) \right] - T_\DD - V_\DD } \: .
\end{equation}
Especially, this propagator includes the residual interaction $V_\DD$ 
in the excited $\DD$ system, see also fig.\ \ref{fig:potential}(c).
$V_\DD$, as well as the transition potential $V_{\DD \to NN}$ between
the intermediate $\DD$ system and the $NN$ state, are both constructed
within the same meson exchange model as $V_\DN$. 
To the $\DD \to NN$ transition potential, of course, only the $\pi$
and $\rho$ meson exchanges contribute. The $\DD$ excitation can thus
be separated into a spin--longitudinal part (with $\pi$--like 
coupling to $NN$) and a spin--transverse part (with $\rho$--like
coupling to $NN$). As we will demonstrate later, this spin--structure
of the $\DD$ excitation is important for the angular distribution of the pions.

The matrix element corresponding to the second mechanism where only one
$\Delta$ is excited, fig.\ \ref{fig:diagrams}(b), is found to be
\begin{equation}
  \label{eq:10}
  {\mathcal M}_\DN = \: \langle \psi_n | \: \langle \psi_p | 
  \: \tilde F^\dagger_{\pi\pi} (\vec k_a, \vec k_b \,) 
  \: | \psi_d \rangle
\end{equation}
Here, the operator $\tilde F^\dagger_{\pi\pi}$ is obtained from 
$F^\dagger_{\pi\pi}$ by simply replacing for the second pion 
$\vec S^\dagger$ and $\vec T^\dagger$ with their hermitian adjungates 
$\vec S$ and $\vec T$, respectively.

In order to obtain the $2\pi$ production amplitude, the coherent 
sum of the two contributions of eqs. (\ref{eq:9}, \ref{eq:10}) has to be 
taken. Due to the fact that the residual interactions $V_\DN$ and $V_\DD$
couple states of different quantum numbers, the concrete evaluation
of the matrix elements leads to a system of coupled integro--differential
equations. It turns out that they can be solved in a very effective
way with the so--called Lanczos method \cite{whitehead77}. This
procedure has already been described in ref.\ \cite{udagawa94} 
and can easily be applied to the present calculation.


\section{Results and Discussion}
\label{sec:2}

\subsection{Parameters of the model}

Input parameters of the model are the meson and baryon masses, coupling 
constants, and formfactor cutoffs at each vertex. The values we used in our 
calculations are given in table \ref{tab:parameters}. 
Most of these quantities, of course, are physical observables that 
are already determined from other experiments, e.g. from $NN$ scattering
\cite{dumbrajs83}. 
Only the sigma meson and cutoff parameters remain to be fixed.
They have been adjusted to fit the experimental data for the one pion 
production $N+N \to ^2$H$ \, \pi$. 
We made sure that not only the total cross section but also
the angular distributions and analyzing powers are correctly reproduced 
in the $\Delta$ resonance energy region \cite{mosbacher98a}. After this consistency check, 
we may now continue and discuss our results for the $2\pi$ production 
$n+p \to ^2$H$\, (\pi\pi)$.

\subsection{The deuteron spectrum}
\label{sec:spectrum}

\begin{figure}
  \begin{center}
    \epsfig{file=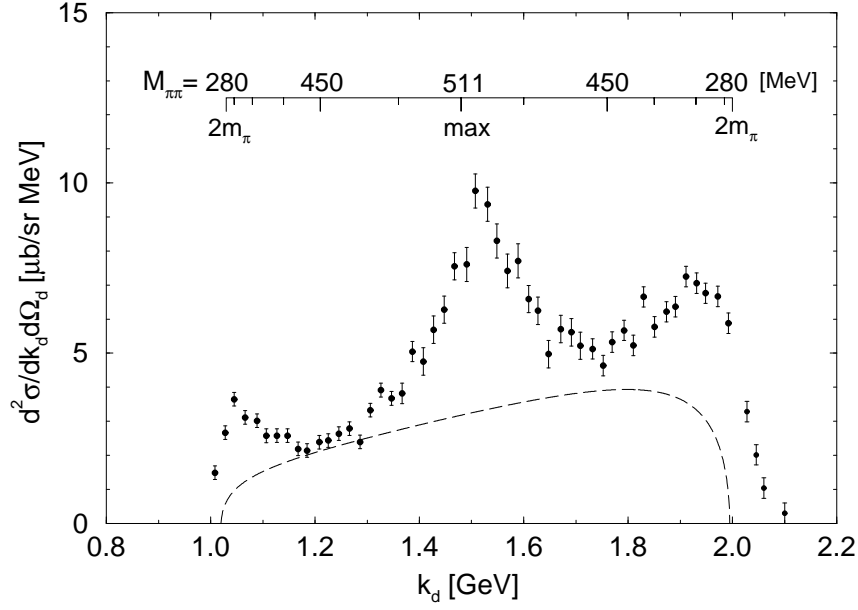, height=8cm}
  \end{center}
  \caption{Experimental deuteron spectrum $d^2 \sigma / dk_d d\Omega_d$
    of the reaction $n+p \to ^2$H$\, (\pi\pi)$, measured at $k_n = 1.88$ GeV
    and $\theta_d = 0^o$. Data have been taken from \protect \cite{plouin78}.
    The dashed line is the phase space for the $2 \pi$ production 
    (in arbitrary units), and $M_{\pi\pi}$ is the invariant mass of the 
    $2 \pi$ system.}
  \label{fig:expspect}
\end{figure}

In fig.\ \ref{fig:expspect}, we present the experimental deuteron spectrum 
$d^2 \sigma / dk_d d\Omega_d$ as it is measured in the reaction 
$n+p \rightarrow ^2H (\pi \pi)$ at a beam momentum $k_n = 1.88$ GeV and 
in forward direction $\theta_d = 0^o$. 
The production of both neutral ($\pi^0 \pi^0$) and charged ($\pi^+ \pi^-$) 
pion pairs is possible. The available phase space for the $2\pi$
production is indicated by the dashed line. 

Obviously, the cross section does not follow the phase space but
shows a characteristic structure which three pronounced peaks.
The two outer maxima correspond to a kinematical situation
where the invariant mass $M_{\pi\pi}$ of the $2\pi$ system is minimal,
i.e. where $M_{\pi\pi} = 2 m_\pi$. The broad central maximum, on the other hand,
is located at the  position where $M_{\pi \pi}$ has its maximal 
value (511 MeV in the present case). In the CMS, this situation will 
be realized if all the kinetic energy is taken by the two pions 
and the deuteron remains at rest.

\begin{figure}
  \begin{center}    
    \epsfig{file=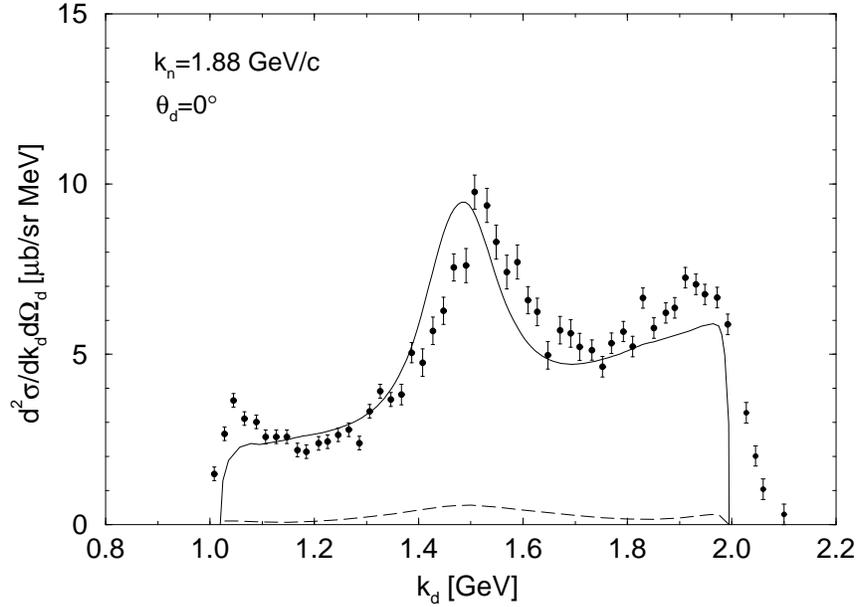, height=8cm}
  \end{center}
  \caption{Theoretical result in comparison with the experimental spectrum
    of fig.\ \protect \ref{fig:expspect}. Solid line: full calculation
    with $\DD$ and $\DN$ mechanism included; dashed line: only $\DN$ excitations.
    (see also fig.\ \protect \ref{fig:diagrams}).}
  \label{fig:theospect}
\end{figure}

As can be seen from fig.\ \ref{fig:theospect}, our model is able to 
explain the characteristic structure of the experimental spectrum. 
The solid line represents the full result with
all intermediate interactions included. The $\DD$ mechanism, for which
the Feynman diagram was presented in fig.\ \ref{fig:diagrams}(a), is in fact
the only important contribution to $2\pi$ production in the energy range under
consideration. The cross section contribution due to the $\DN$ mechanism
of fig.\ \ref{fig:diagrams}(b) is almost negligible. Since the optimal energy
for this second mechanism would be $\sqrt{s} = M_\Delta + M - m_\pi$,
its suppression in the ABC energy region is not surprising.

\begin{figure}
  \begin{center}
    \epsfig{file=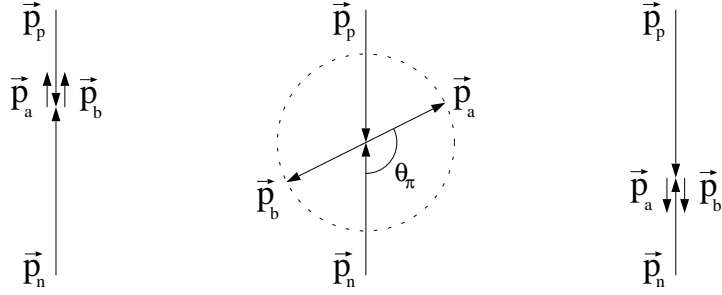, height=4cm}
  \end{center}
  \caption{Kinematical configurations in the deuteron rest frame 
    $\vec p_d = \vec 0$. Right and left: Momentum configuration for the
    outer maxima of the deuteron spectrum; middle: momentum configuration
    for the central maximum.}
  \label{fig:kinematics}
\end{figure}

With focus on the $\DD$ excitation and following the arguments of ref.\ 
\cite{risser73}, the origin of the different maxima can
be understood quite easily. Let us define
\begin{equation}
  \label{eq:res7}
  \mathbf K = \mathbf p_a + \mathbf p_b \quad \text{and} \quad 
  \mathbf k = \mathbf p_a - \mathbf p_b \, ,
\end{equation}
where $\mathbf p_a, \mathbf p_b$ denote the four--momenta of the 
two pions. The excitation of a $\DD$ system is most effective if
the invariant masses of the two $\Delta$'s are approximately equal
(and hence both close to the resonant mass). In this case we have
\begin{equation}
  \label{eq:res8}
  s_{\Delta_a} = \frac{1}{4} \: ({\mathbf p_d + \mathbf K + \mathbf k})^2 = 
  \frac{1}{4} \: ({\mathbf p_d + \mathbf K - \mathbf k})^2 = s_{\Delta_b} \: ,
\end{equation}
and since $\mathbf K \cdot \mathbf k = 0$ this is equivalent to 
\begin{equation}
  \mathbf p_d \cdot \mathbf k = 0 \, .
\end{equation}
This condition can be fullfilled in two ways:
\begin{enumerate}
\item  $\mathbf k = 0$ and therefore $M_{\pi\pi} = \sqrt{\mathbf K^2} = 2 m_\pi \:$, or
\item $\vec p_d = 0$ in the CMS of the two pions (which means that the deuteron restframe
  and the $2\pi$ restframe are identical). In the latter case, the two pions carry
  the whole kinetic energy, and $M_{\pi\pi} = $ max.
\end{enumerate}
The first situation leads to the outer maxima and corresponds to the parallel
decay of the two $\Delta$'s because of $\mathbf p_a \approx \mathbf p_b$, i.e. 
the relative momentum is very small.
The second situation explains the central maximum and corresponds 
to the antiparallel decay of the $\DD$ excitation for which the relative
momentum is maximal.  
In the deuteron restframe, the different kinematical configurations for the 
three maxima can be visualized as depicted in fig.\ \ref{fig:kinematics}.

\subsection{Angular distribution of the pions}
\label{sec:angdist}

\begin{figure}
  \begin{center}
    \epsfig{file=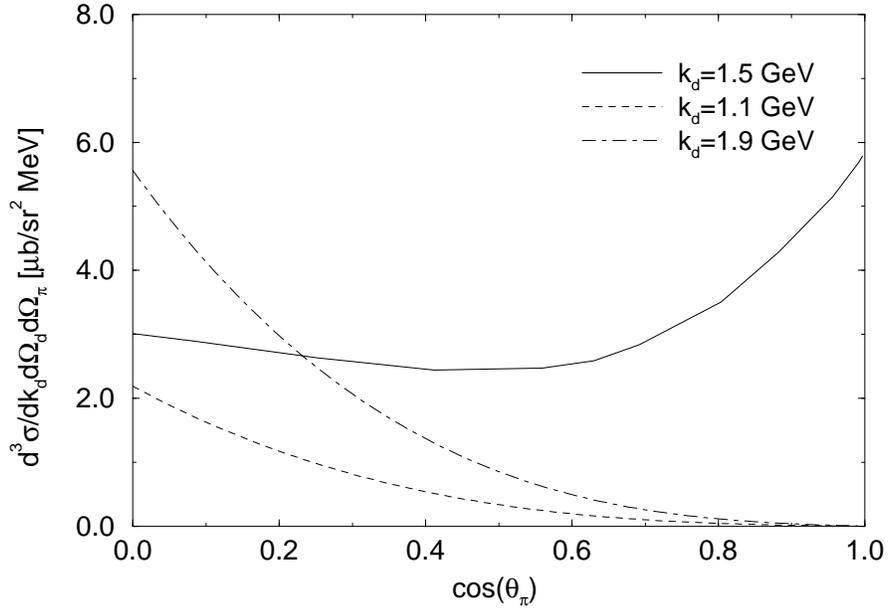, height=8cm}
  \end{center}
  \caption{Angular distribution $d^3 \sigma / dk_d d\Omega_d d\Omega_\pi$
     of the pions, for $k_n = 1.88$ GeV and $\theta_d = 0^o$. 
     $\theta_\pi$ is the angle between the relative pion momentum
     and the beam axis in the deuteron rest frame. The different
     deuteron laboratory momenta of 1.1, 1.5 and 1.9 GeV correspond
     to the left, central, and right maximum of the deuteron spectrum.}
  \label{fig:angdist1}
\end{figure}

The kinematical situations leading to the three maxima of the deuteron spectrum
are also clearly visible in the angular distribution of the pions.
In fig.\ \ref{fig:angdist1}, we show the triple differential cross section
$d^3 \sigma / dk_d d\Omega_d d\Omega_\pi$ for different deuteron laboratory momenta.
It is plotted as a function of $\theta_\pi$ which is the angle between the
relative momentum $\vec k$ of the two pions and the beam axis, as given 
in the deuteron rest frame (see also fig.\ \ref{fig:kinematics}).

For the two outer maxima ($k_d^{\text{lab}} = 1.1$ GeV and 
$k_d^{\text{lab}} = 1.9$ GeV), the pion momenta
are nearly parallel and hence the relative momentum is dominantly perpendicular
to the beam axis. Consequently, the differential cross section reaches 
its largest value at $\cos \theta_\pi = 0$.

For the central maximum  ($k^{\text{lab}}_d = 1.5$ GeV), however, the pion momenta are 
antiparallel, and all angles $\theta_\pi$ are kinematically possible. 
If there was no spin dependence of the excitation and the residual interaction,
an isotropic distribution would be the result. The observed angular variation
of the cross section is thus reflecting the spin structure of the process
which mainly follows from the $\pi$ and $\rho$ exchange contributions to 
the $NN \! \rightarrow \DD$ transition potential.

In order to examine this spin structure in more detail, we will neglect 
for the moment the residual interaction and the possibility of spin--flips.
Then, the spin--longitudinal $\pi$ exchange leads to an operator 
proportional to
\begin{equation}
  \left[ (\vec S_1 \cdot \vec q \,) \, (\vec S_1^\dagger \cdot \vec k_a) \right] \;
  \left[ (\vec S_2 \cdot \vec q \,) \, (\vec S_2^\dagger \cdot \vec k_b) \right] 
  \, \sim \, (\vec q \cdot \vec k_a \,) \, (\vec q \cdot \vec k_b \,) \, ,
\end{equation}
where $\vec k_a, \vec k_b$ are the pion momenta and $\vec q$ is the momentum
transfer from the neutron projectile to the proton target. If both the pion momenta
are parallel (or antiparallel) to $\vec q$, the corresponding cross section
will be largest. This results in a maximum at $\cos \theta_\pi = 1$ in the
spin--longitudinal channel.

In complete analogy, the spin--transverse $\rho$ exchange has the operator structure
\begin{equation}
  \left[ (\vec S_1 \times \vec q \,) \, (\vec S_1^\dagger \cdot \vec k_a \right] \;
  \left[ (\vec S_2 \times \vec q \,) \, (\vec S_2^\dagger \cdot \vec k_b) \right]
  \, \sim \, (\vec q \times \vec k_a) \, (\vec q \times \vec k_b) \, .
\end{equation}
Therefore, the maximum in the spin--transverse channel is reached if both the pion momenta
are perpendicular to $\vec q$, i.e. if $\cos \theta_\pi = 0$.

\begin{figure}
  \begin{center}
    \epsfig{file=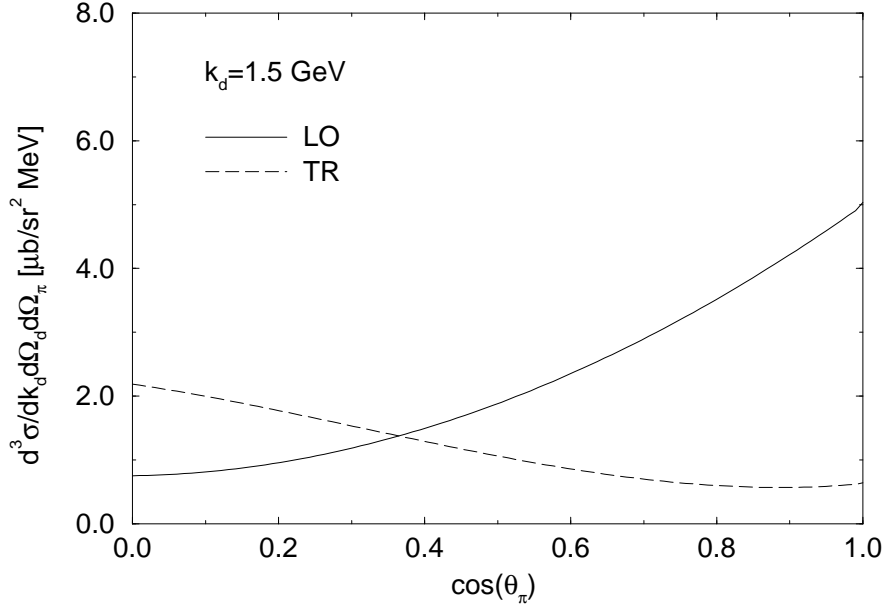, height=8cm}
  \end{center}
  \caption{Contributions of the spin--longitudinal (LO) and 
    spin--transverse (TR) component of the transition potential 
    $NN \to \DD$ to the differential cross section at $k_d = 1.5$ GeV
    (without residual interaction).}
  \label{fig:angdist2}
\end{figure}

Indeed, the theoretical curves in fig.\ \ref{fig:angdist2} exactly reflect
the expected shapes. We conclude that a measurement of the pion angular 
distribution would be very helpful in order to reveal the spin structure 
of the interaction.

\subsection{Influence of the residual interaction}
\label{sec:vinfl}

\begin{figure}
  \begin{center}  
    \epsfig{file=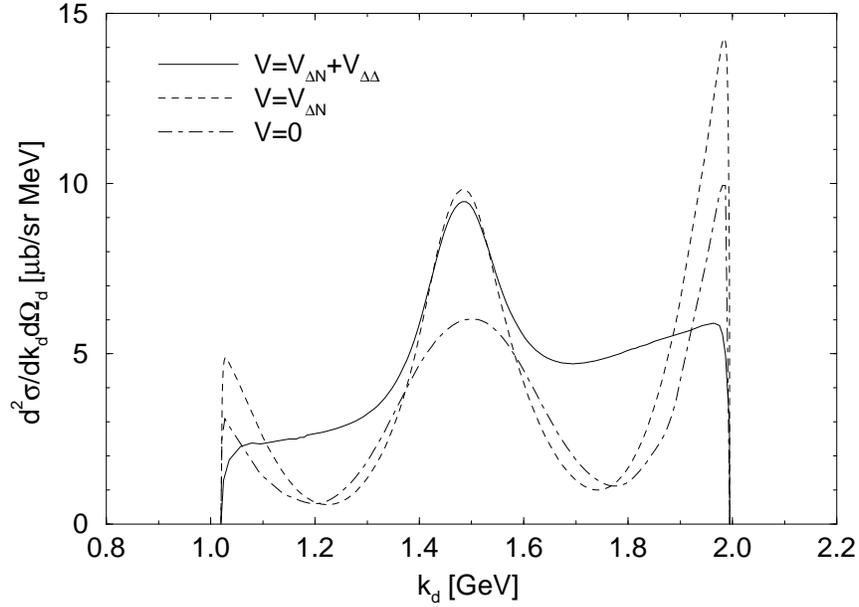, height=8cm}
  \end{center}
  \caption{Influence of the residual interaction on the deuteron spectrum,
    for $k_n = 1.88$ GeV and $\theta_d = 0^o$. Solid line: full calculation
    with both $\DN$ and $\DD$ potential; dashed: with $\DN$ potential but 
    without $\DD$ potential; dash--dotted: without any residual interaction.}
  \label{fig:vinfl}
\end{figure}

Effects of residual interactions can be examined in both the $\DD$ and the 
$\DN$ system since these are the intermediate configurations in our model.
The influence of the corresponding interaction potentials on the theoretical
spectrum is shown in fig.\ \ref{fig:vinfl} for the case of forward scattering
($\theta_d = 0^o$).

The dash--dotted line was calculated without any residual interaction, 
i.e. with $V=0$ for the intermediate states. The typical ABC structure with 
its three peaks as discussed in section \ref{sec:spectrum} is clearly present.   
Our result comes quite close to the fully relativistic calculation of 
Bar--Nir et al. \cite{barnir75} which also does not include the residual
interactions.

By choosing $V=V_{\DN}$ but still neglecting the direct $\DD$ interaction,
we obtain the dashed line. Here, the maxima are even more pronounced.
The enlargement of the peak cross sections is due to the reduction of the
excitation energy for the $\DN$ system which is caused by the attractive
$\DN$ potential. For $\sqrt{s} < 2 M_\Delta$ as in the present case,
the whole system is thus closer to the resonance energy. The relative 
position of the two invariant $\Delta$ masses remains however unchanged
since in our model they are fixed independently of $V_{\DN}$.

On the other hand, the $\DD$ potential can directly influence the relative 
wave function of the $\DD$ system via a redistribution of momentum
within the two particle system.
The solid line in fig.\ \ref{fig:vinfl} has been calculated with the full 
residual interaction $V = V_\DN + V_\DD$. The inclusion of $V_\DD$ results in
a transfer of strength from the maxima to the regions between which are 
kinematically less favoured. This transfer takes place because the optimal configuration 
of equal $\Delta$ masses can now be reached even if the initial energy distribution
of the two pions was asymmetric. As a consequence, the cross section is enhanced 
between the peaks and simultanously the magnitude of the peaks is reduced. 
This reduction is more prominent for the outer maxima since they are related 
to a small relative momentum of the $\DD$ system. For the central maximum,
the relative momentum is large and therefore only a slight influence 
of the $\DD$ potential is observed.

Obviously, the residual interactions play quite an important role in 
the $2\pi$ production and significantly influence the deuteron spectrum.
The inclusion of these effects in the calculation clearly improves the agreement 
between theory and experiment.

\subsection{Angular and energy dependence of the deuteron spectrum 
  and the total cross section}

\begin{figure}
  \begin{center}
    \epsfig{file=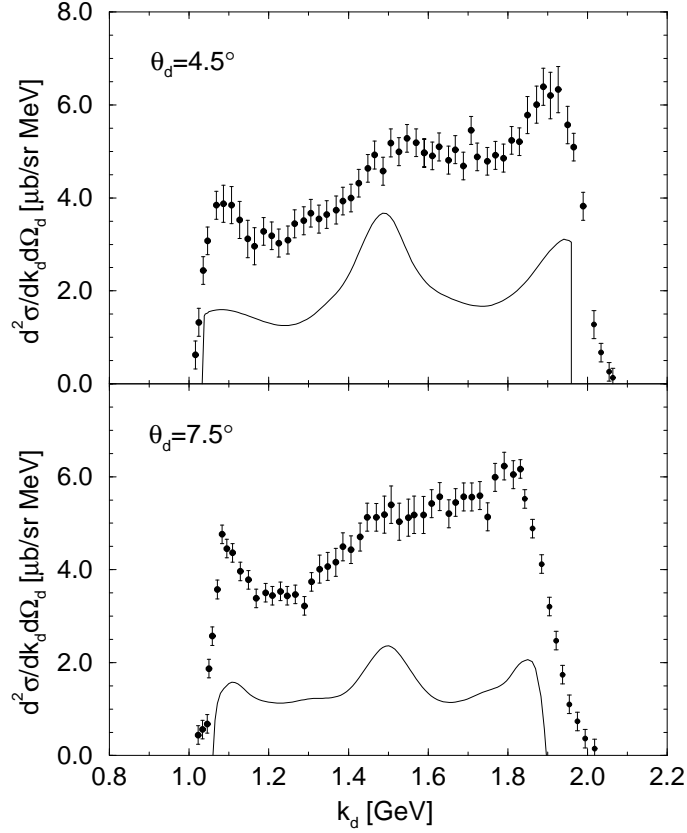, height=11cm}
  \end{center}
  \caption{Deuteron spectra in the reaction $n+p \to ^2$H$\, (\pi\pi)$ at
    $k_n = 1.88$ GeV and $\theta_d = 4.5^0$ and $7.5^o$, respectively.
    The solid line is the theoretical result. Experimental data from ref.\ 
    \protect \cite{plouin78}.}
  \label{fig:anguldep}
\end{figure}

Studying the dependence of the deuteron spectrum on the scattering angle
$\theta_d$, we find that the theoretical cross section for $\DD$ excitation
is decreasing too fast. Exemplatory we present in fig.\ \ref{fig:anguldep}
our calculations for $\theta_d = 4.5^o$ and $\theta_d = 7.5^o$. The experimental 
data \cite{plouin78} are underestimated by a factor $2 \sim 3$. 
Other microscopic calculations, e.g. assuming a two nucleon exchange as the
dominant reaction mechanism, yield a comparable angular dependence and are
also not able to describe the data \cite{Anjos73,Anjos76}. The reason for this behaviour is not
understood. Maybe more than only one production mechanism has to be taken into
account in order to solve this problem.

\begin{figure}
  \begin{center}  
    \epsfig{file=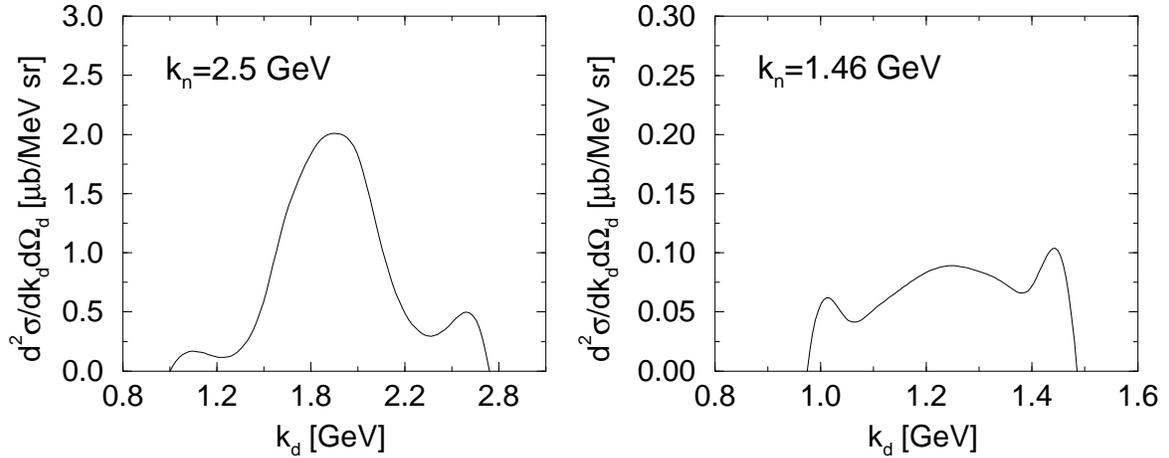, height=6cm}
  \end{center}
  \caption{Theoretical deuteron spectra in our model for forward 
    scattering $\theta_d = 0^o$ at beam momenta $k_n = 2.5$ GeV and 
    $k_n = 1.46$ GeV, respectively.}
  \label{fig:energdep}
\end{figure}

Fig. \ref{fig:energdep} is demonstrating the energy dependence
of the deuteron spectrum in our model. One recognizes that the central maximum 
is getting more pronounced if the total energy increases. 
We remind the reader that this peak corresponds to the $\DD$ excitation 
with maximal relative momentum, which already gives the natural explanation 
of the observed effect. 

\begin{figure}
  \begin{center}
    \epsfig{file=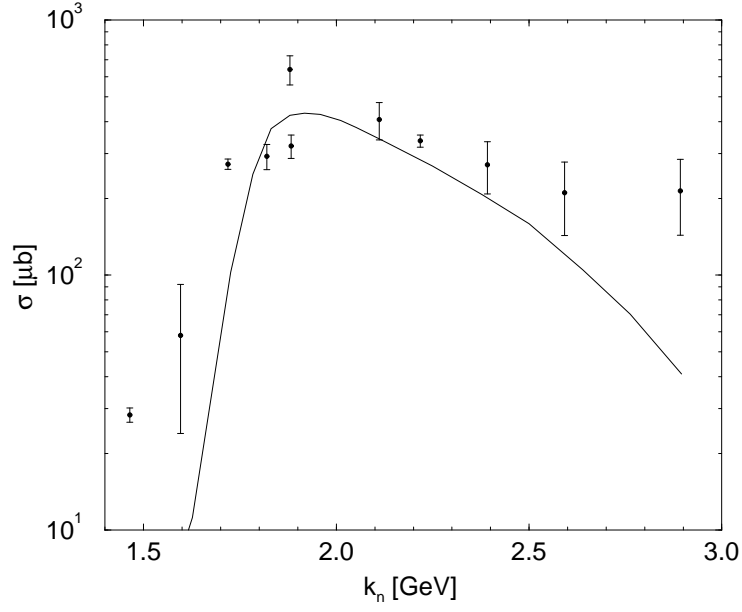, height=8cm}
  \end{center}
  \caption{Total cross section for $n+p \to ^2$H$\, (\pi\pi)$ as 
     a function of the neutron momentum. Experimental data from 
     \protect \cite{plouin78,abdivaliev80,barnir73,hollas82}.
     For $k_n = 2.09$ GeV, we have $\sqrt{s} = 2 M_\Delta$.
     The threshold for $2 \pi$ production is at $k_n = 1.19$ GeV.}
  \label{fig:totcs}
\end{figure}

After integration over deuteron momentum and scattering angle, the total cross 
section for the $n+p \rightarrow ^2H (\pi\pi)$ reaction is obtained.
In fig.\ \ref{fig:totcs}, our theoretical result is compared to the 
experimental data \cite{plouin78,abdivaliev80,barnir73,hollas82}.
In the $\DD$ resonance energy region, i.e. for a neutron momentum 
of $k_n \approx 2$ GeV, the agreement is quite good. This confirms our 
assumption that the $2\pi$ production in the regime of the ABC effect
is dominated by the $\DD$ excitation mechanism. On the other hand, 
one recognizes that the experimental cross section close to 
the $2\pi$ production threshold ($k_n = 1.19$ GeV) is understimated by 
nearly two orders of magnitude. In this case,
the total energy of $\sqrt{s} = M_d + 2 m_\pi \ll 2M_\Delta$    
is simply not sufficient for the excitation of a $\DD$ system,
and other production mechanisms will be more important.
To mention just two possibilities, the pions could be produced 
by s--wave rescattering or via excitation of the $N^* (1440)$ resonance,
as recently discussed in \cite{alvarez98}. Since there is also
experimental evidence \cite{hollas82} that the ABC structure completely 
disappears when approaching the $2 \pi$ production threshold, further
theoretical investigations of the low energy region would be of
high interest.

\section{Summary and Conclusions}
\label{sec:3}

To summarize we have shown that the $\Delta \Delta$ excitation is the 
dominant reaction mechanism in the $n+p \to ^2$H$\, (\pi\pi)$ two pion 
production at $k_n = 1.88$ GeV/c. It is able to explain the ABC effect
as observed in the experimental deuteron spectrum. Hereby, the residual 
interactions in the intermediate $\DD$ and $\DN$ states play an important role.
We obtain good agreement with experimental data at forward scattering
but too fast a decrease of the cross section for higher scattering
angles of the deuteron. The total cross section is also underestimated
close to the $2 \pi$ production threshold where other reaction
mechanisms get more important.
Furthermore, we found that the spin structure of the $\DD$ 
excitation directly influences the angular distribution of the two pions.
We conclude that the $2 \pi$ production in the ABC energy region 
may well serve as a tool for closer examination of direct $\DD$ 
interaction and transition potentials.

\section*{Acknowledgments}

This paper is dedicated to the memory of Prof. Dr. Franz Osterfeld
who deceased in 1997. 
We would like to thank J. Speth, Ch. Hanhart, O. Krehl and C. Wilkin 
for many helpful discussions. The work was supported in part by the 
Studienstiftung des deutschen Volkes.



\newpage

\begin{table}
  \caption{Parameters of the model.}
  \begin{tabular}{cccccc}
    & $f^2_{\alpha NN} /4\pi$ & $f^2_{\alpha N \Delta} /4\pi$ 
    & $f^2_{\alpha \Delta \Delta} /4\pi$ & $\Lambda_\alpha$ [GeV] 
    & $m_\alpha$ [MeV] \\
    \tableline
    $\pi$ & 0.081 & 0.32 & 0.0031 & 1.1 & 138 \\
    $\rho$ & 5.4 & 21.6 & 0.286 & 1.4 & 770 \\
    $\omega$ & 8.1 \tablenotemark[1] & --- & 8.1 \tablenotemark[1] 
    & 1.7 & 783 \\
    $\sigma$ & 6.9 \tablenotemark[1] & --- & 6.9 \tablenotemark[1] 
    & 1.6 & 580 \\
  \end{tabular}
  \tablenotetext[1]{$g^2_{\alpha NN}/4\pi$ resp.\ 
    $g^2_{\alpha \Delta \Delta}/4\pi$ is given.}
\label{tab:parameters}
\end{table}

\pagebreak

\end{document}